\documentclass[9pt, conference]{IEEEtran}
\IEEEoverridecommandlockouts
\usepackage{cite}
\usepackage{amsmath,amssymb,amsfonts}
\usepackage{algorithmic}
\usepackage{graphicx}
\usepackage{textcomp}
\usepackage{xcolor}
\usepackage{bm}
\usepackage{upgreek}
\usepackage{color}
\usepackage{tikz}
\usepackage{pgfplots}
\usepackage{scalerel}
\usepackage{multirow}
\usepackage{makecell}
\usepackage{booktabs}
\usepackage{mathtools}
\usepackage[caption=false]{subfig}
\pgfplotsset{compat=1.7}
\def\BibTeX{{\rm B\kern-.05em{\sc i\kern-.025em b}\kern-.08em
    T\kern-.1667em\lower.7ex\hbox{E}\kern-.125emX}}

\def\BibTeX{{\rm B\kern-.05em{\sc i\kern-.025em b}\kern-.08em
    T\kern-.1667em\lower.7ex\hbox{E}\kern-.125emX}}

\begin{document}
%topmargin=0mm
\title{Domain-Incremental Learning for Audio Classification
\thanks{This work was supported by Jane and Aatos Erkko Foundation under grant number 230048, "Continual learning of sounds with deep neural networks". \\
The authors wish to thank CSC-IT Centre of Science Ltd., Finland,  for providing computational resources.} 
}

\author{
\IEEEauthorblockN{Manjunath Mulimani, Annamaria Mesaros}
\IEEEauthorblockA{\textit{Signal Processing Research Centre}, \textit{Tampere University}, Tampere, Finland \\
\{manjunath.mulimani, annamaria.mesaros\}@tuni.fi
}}
\makeatletter
\def\ps@IEEEtitlepagestyle{%
  \def\@oddfoot{\mycopyrightnotice}%
  \def\@oddhead{\hbox{}\@IEEEheaderstyle\leftmark\hfil\thepage}\relax
  \def\@evenhead{\@IEEEheaderstyle\thepage\hfil\leftmark\hbox{}}\relax
  \def\@evenfoot{}%
}
\def\mycopyrightnotice{%
  \begin{minipage}{\textwidth}
  \centering \scriptsize
  Copyright~\copyright~2025 IEEE.  Personal use of this material is permitted. Permission from IEEE must be obtained for all other uses, in any current or future media, including reprinting/republishing this material for advertising or promotional purposes, creating new collective works, for resale or redistribution to servers or lists, or reuse of any copyrighted component of this work in other works.
  \end{minipage}
}
\maketitle

\begin{abstract}
In this work, we propose a method for domain-incremental learning for audio classification from a sequence of datasets recorded in different acoustic conditions. Fine-tuning a model on a sequence of evolving domains or datasets leads to forgetting of previously learned knowledge. On the other hand, freezing all the layers of the model leads to the model not adapting to the new domain. In this work, our novel dynamic network architecture keeps the shared homogeneous acoustic characteristics of domains, and learns the domain-specific acoustic characteristics in incremental steps. Our approach achieves a good balance between retaining the knowledge of previously learned domains and acquiring the knowledge of the new domain. We demonstrate the effectiveness of the proposed method on incremental learning of single-label classification of acoustic scenes from European cities and Korea, and multi-label classification of audio recordings from Audioset and FSD50K datasets. The proposed approach learns to classify acoustic scenes incrementally with an average accuracy of 71.9\% for the order: European cities $\rightarrow$ Korea, and 83.4\%  for Korea $\rightarrow$ European cities. In a multi-label audio classification setup, it achieves an average $l\omega lrap$ of 47.5\% for Audioset $\rightarrow$ FSD50K and 40.7\%  for FSD50K $\rightarrow$ Audioset.
\end{abstract}


\section{Introduction}
Domain-incremental learning  (DIL) can be described as the ability of a model to continuously acquire new knowledge over time in ever-changing environments without forgetting the previously learned knowledge.
Most of the audio research today focuses on developing  models to solve various machine listening tasks such as acoustic scene classification (ASC) \cite{mesaros2018acoustic}, multi-label audio classification \cite{Fonseca2018_DCASE, kong2020panns}, sound event detection (SED) \cite{mesaros2021sound}, etc., and these models are specialized to a specific dataset or environment or domain. The use of such models to continuously or incrementally learn new audio domains leads to catastrophic forgetting of previously learned domains. Specifically, the new knowledge overwrites the previous knowledge in the absence data of previous domains. Shifting from one domain to another domain, i.e., domain shift, can cause catastrophic forgetting due to changes in background locations, recording devices or any other acoustic conditions. 

The naive way to avoid forgetting is to train a separate model for each domain, or to store the data from all domains and retrain a single model whenever a new domain arrives. Both these solutions are computationally expensive and require significant storage to save all the data and models, which may not be possible in some application due to privacy and hardware constraints.

In this work, we aim to develop a universal DIL system that learns to classify audio from  different domains sequentially over time without forgetting the knowledge of any of the previously learned domains. DIL was successfully applied to detect objects on the road from different visual geographical locations \cite{garg2022multi} and weather conditions \cite{mirza2022efficient}, and to classify acoustic scenes from audio recorded in different geographical locations \cite{mulimani2024online}.  

DIL is different from existing domain adaptation (DA) methods used for ASC from different devices \cite{gharib2018unsupervised, drossos2019unsupervised, mezza2021unsupervised}. DA setup typically includes two domains: source and target. It transfers the knowledge from the source to the target domain and only focuses on the accuracy in the target domain.  DA requires access to the data of the source domain to match the distribution with the target domain. In comparison to DA, the DIL setup can have multiple domains to adapt over time, it focuses on the overall accuracy in all the domains seen so far, takes additional measures to alleviate the forgetting, and, typically, does not have access to the previous domain's data. 
DIL is also different from existing class-incremental learning (CIL) methods \cite{wang2021few, mulimani2024class} where the model learns new classes incrementally over time, typically from the same domain. In contrast, DIL learns the same classes from continuously evolving domains.

In our previous work, we proposed a DIL approach for ASC with data from different locations \cite{mulimani2024online}. Specifically, during training we only updated and stored the domain-specific statistics, i.e., running mean and variance, with all other parameters kept frozen. During inference, the domain-id along with the test sample is provided to the classifier to choose the corresponding  statistics to classify the sample. This method does not forget any of the previous domains and achieves maximum stability, i.e., ability of the model to retain the existing knowledge of already learned domains. However, only statistics did not achieve better performance on the new domain as compared to fine-tuning method given in \cite{mulimani2024online}, hence it exhibited lower plasticity, i.e., ability of the model to acquire knowledge of new  domains.

In this work, we propose a dynamic framework that reparameterizes the network architecture into domain-shared and domain-specific parameters for DIL of audio tasks. Domain-shared parameters are universal and shared by all domains. We only update the domain-specific parameters to learn a new domain at each incremental time step.  This construction allows the model to achieve a good stability-plasticity trade-off.

 The contributions of this work are as follows:
 %\begin{itemize}
(1) we investigate the performance of the proposed method in both single-label and multi-label audio classification;
(2) we propose a domain-agnostic approach that classifies the audio by automatically identifying the domain-specific parameters, and compare its performance with domain-aware setup above;
(3) we investigate the performance of the proposed approach in all possible orders of the input domains.
% \end{itemize}

The rest of the paper is organized as follows: Section 2 presents the notations and the proposed DIL method for audio classification. Section 3 introduces the datasets, baselines, implementation details, and results. Finally, conclusions are given in Section 4.

\section{Domain-Incremental Learning}
\subsection{DIL setup and notations}

In our DIL setup, we present a sequence of audio classification tasks to a model; these tasks represent the different datasets of domains: $\mathcal{D}_1, \mathcal{D}_2, ..., \mathcal{D}_t, ..., \mathcal{D}_T$. The existing model learns current task, i.e., $\mathcal{D}_t$ in our case, at incremental time step $t$. A domain $\mathcal{D}_t$ is an audio dataset collected in different acoustic conditions than the previous domains, composed of audio clips and corresponding class labels. All domains share the same classes. We aim to train a single audio classification model that learns to classify the same data when domain or data distribution changes sequentially.  More importantly, the performance of the model should not degrade on any of the previous domains $\mathcal{D}_{t-i}, 0<i<t$  when it learns a new domain $\mathcal{D}_t$, given that data of any previous domains $\mathcal{D}_{t-i}$ is not available at any step $t$. Note that in this work we refer to $\mathcal{D}_t$ as task, dataset and domain interchangeably.

\subsection{DIL method for Audio Classification}
\begin{figure}[!tbp]
  \centering
  \includegraphics[width=0.4\linewidth]{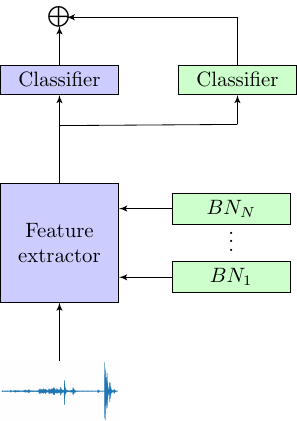}  
  \caption{Overview of the proposed Domain-Incremental Learning approach at an incremental time step. The layers in \textcolor{green}{green} are domain-specific. The classifier in \textcolor{blue}{blue} is domain-shared. Feature extractor includes the domain-specific BN layers of previous domain and the domain-shared CNN layers. }
     \label{fig:domain}
     \vspace{-10pt}
\end{figure}

An overview of the proposed method is given in Fig.~\ref{fig:domain}. The system is composed of shared and domain-specific layers. 
Domain-specific layers are added to the common base model trained on domain $\mathcal{D}_1$.
The idea is to factorize the model latent space so that homogeneous acoustic characteristics of the domains are captured in the domain-shared parameters $\mathcal{W}_b$ and remains unchanged. On the other hand, heterogeneous acoustic characteristics of the domains are learned by the domain-specific parameters $\mathcal{W}_t$, which are exclusive to a specific domain $\mathcal{D}_t$.  For instance, a \textit{car horn} may have some common characteristics irrespective of location, captured in  $\mathcal{W}_b$, and  location-specific characteristics (e.g. Europe/Korea), captured in $\mathcal{W}_t$. 

\begin{figure}[!tbp]
  \centering
  \includegraphics[width=0.4\linewidth]{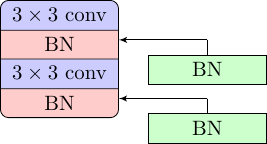}  
  \caption{A CNN block at incremental time step. The layers in \textcolor{blue} {blue} and \textcolor{green}{green} are domain-shared and domain-specific respectively. The layers in \textcolor{red}{red} are domain-specific layers of previous domain.}
     \label{fig:cnn_block}
     \vspace{-10pt}
\end{figure}

Our approach is designed for CNN-based models typically used for audio classification.
Traditional $3 \times 3$ convolutional layers followed by a classifier or output layer,  of a base model are domain-shared layers  $\mathcal{W}_b$. Domain-specific layers are the BN and output layers  $\mathcal{W}_t$ of each incremental domain. For instance, in a traditional convolutional block, shown in Fig.~\ref{fig:cnn_block}, the convolutional layers are shared among all domains and BN layers are exclusive to a specific domain.

BN layers normalize the input $\bm{h}$ into $\hat{\bm{h}} = (\bm{h}-\bm{\mu})/\sqrt{\bm{\sigma}^2}$ using its $\bm{\mu}$ and standard deviation $\bm{\sigma}$. Further, $\hat{\bm{h}}$ is transferred into $\bar{\bm{h}}=\gamma\hat{\bm{h}}+\beta$ using affine scale $\gamma$ and shift $\beta$ parameters. Specifically, the statistics $\bm{\mu}$ and $\bm{\sigma}$ are computed from input data, and $\gamma$ and $\beta$ parameters are optimized by the loss function during training.
Statistics computed over training data fail to perform well on test data if the distribution of the  test data is significantly different.
In our previous work \cite{mulimani2024online}, we proposed updating the statistics using data of each incremental domain. In this work, we both update the statistics and optimize the transformation parameters for each domain. 
Each domain has its own classifier $\mathcal{G}_t$ whose output is added to the output of the base classifier $\mathcal{G}_1$ through residual connection. The optimization of only the transformation parameters of BN layers and parameters of classifier i.e., $\mathcal{W}_t=\{\gamma, \beta, \mathcal{G}_t\}$, for each domain 
is computationally inexpensive. 

\textbf{Training phase:} In each incremental time step $t$, to learn a new domain, we only train the domain-specific parameters $\mathcal{W}_t$ on $\mathcal{D}_t$ using specific loss for single and multi-label classification, depending on the experimental case. All other parameters, the domain shared $\mathcal{W}_b$ and domain-specific parameters of previous domains remain frozen.

\textbf{Inference phase:} Performance of the model is evaluated on the domains seen so far in two scenarios: domain-aware and domain-agnostic. In a domain-aware setup, input to the model is a combination of domain-id and test sample. Domain-id identifies the layers of the corresponding domain before classifying the test sample. In a domain-agnostic setup, we predict the domain-specific layers to be used with domain-shared layers using uncertainty in the model predictions. Specifically, we forward pass the input through a combination of shared and domain-specific layers of each domain seen so far and obtain the probabilities from the layers of each current domain $\mathcal{D}_t$. Subsequently, we compute the uncertainty $\mathcal{U}(\mathcal{D}_t)$ on given input $x$ among the predicted probabilities $p(y^{\mathcal{D}_t}_c|x)$ using entropy:
\begin{equation}
    \mathcal{U}(\mathcal{D}_t) = -\sum_{c=1}^{C} p(y^{\mathcal{D}_t}_c|x)\log  p(y^{\mathcal{D}_t}_c|x),
    \label{uncertainity}
\end{equation}
where $y$ is the output from $\mathcal{D}_t$ layers and  $C$ is the number of classes in  $\mathcal{D}_t$. We select the layers of a domain which has minimum entropy, denoting lower uncertainty. Hereafter, we refer to this approach as audio domain-incremental learning (ADIL) approach.

\section{Evaluation and Results}
\subsection{Datasets, training setup and baselines}
\textbf{Single-label acoustic scene classification} uses 5 independent datasets %: $\mathcal{D}_1$ to $\mathcal{D}_5$, 
recorded in different geographical locations:
(1) the TUT Urban Acoustic Scenes 2018 development dataset \cite{Mesaros2018_DCASE}, containing samples from 6 different European cities; (2)-(4) audio samples from  Lisbon, Lyon and Prague from TAU Urban Acoustic Scenes 2019 development dataset; (5) samples from Korea \cite{jeong2022cochlscene}. For complete details about class labels and the number of training/testing samples in each dataset, we refer the reader to \cite{mulimani2024online}. 
Experiments are conducted in 2 orders: (1) 6 European cities $\rightarrow$ Lisbon $\rightarrow$ Lyon $\rightarrow$ Prague $\rightarrow$ Korea (Europe $\rightarrow$ Korea order) (2) Korea  $\rightarrow$ 6 European cities $\rightarrow$ Lisbon $\rightarrow$ Lyon $\rightarrow$ Prague  (Korea $\rightarrow$ Europe order). 
For Europe $\rightarrow$ Korea, we use the same 10 classes present in all European cities, then  4 classes from Korea, which overlap with the European set. For Korea $\rightarrow$ Europe, we select 4 classes from Korea and use the same 4 classes throughout the entire chain.

\textbf{Multi-label audio classification}:
We select the 50 largest classes from  the temporally strong Audioset \cite{hershey2021benefit}, of which 35 classes are also available in the FSD50K dataset \cite{fonseca2021fsd50k}. We conduct the experiments in 2 orders: (1) Audioset (50 classes) $\rightarrow$ FSD50K (35 classes in other domain) and (2) FSD50K (35 classes) $\rightarrow$ Audioset (35 classes in other domain). 
Note that Audioset data is from YouTube and FSD50K is from Freesound.

We compare the proposed approach to a few different methods that can be used to solve the same problem: 
 (1) \textit{Feature extraction (FE)}: the feature extractor component of the model is frozen after learning  $\mathcal{D}_1$, and only the classifier (last layer) is updated in each incremental domain;  
(2) \textit{Fine-tuning (FT)}: a current model is fine-tuned on the new domain at each incremental time step, the model being trained incrementally;
(3) \textit{Single-task}: a separate model is trained on each domain. 
(4) \textit{Multi-task}: a model is trained using all the data of the domains seen so far; this approach violates the DIL setup, but it is tested for completeness.

\subsection{Implementation details and evaluation metrics}
We use 6 convolutional blocks as a feature extractor, with the layer specifications the same as PANNs CNN14 \cite{kong2020panns}.  Global
pooling is applied to the last convolutional layer, to get a fixed-length input feature vector to the classifier. For a fair comparison, the same network is used in all experiments. The base model is trained from scratch on the domain $\mathcal{D}_1$, then adapted to other domains in incremental time steps. Input recordings are resampled to 32 kHz and
log mel spectrograms are computed using default values given in \cite{kong2020panns}. 

The model is optimised using cross-entropy and binary cross-entropy losses for single and multi-label audio classification, respectively. The model is trained using Adam optimizer \cite{loshchilov2017sgdr} with 
a mini-batch size of 32. 
The learning rates to train the model on domain $\mathcal{D}_1$ and on domain $\mathcal{D}_t$ at each incremental time step 
are set to \{0.001 and 0.0001\} following \cite{mulimani2024class}.  
The number of training epochs at each step is set to 120. 
CosineAnnealingLR \cite{loshchilov2017sgdr} scheduler updates the optimizer in every epoch. 

Following the standard practice in incremental learning \cite{mirza2022efficient, mulimani2024online}, we evaluate the performance of the model at each incremental step on the current domain and all previously seen domains using average accuracy and average $l\omega lrap$ \cite{fonseca2019audio}, 
for single and multi-label audio classification respectively.   
Average accuracy / $l\omega lrap$ is the average of accuracy / $l\omega lrap$ values of the method over the current and all previously seen domains.  
In addition, we use forgetting (Fr) to compute the drop in accuracy on previous domains when the model learns a new domain, computed as:
\begin{equation}
    Fr=\frac{1}{t-1}\sum_{i=1}^{t-1} (ACC_{\mathcal{D}_{t-i}, \mathcal{D}_{t-i}} - ACC_{\mathcal{D}_{t}, \mathcal{D}_{t-i}}),
\end{equation}
where $ACC_{\mathcal{D}_{t-i}, \mathcal{D}_{t-i}}$ is the accuracy of the model trained on a previous domain $\mathcal{D}_{t-i}$ and evaluated on the same domain $\mathcal{D}_{t-i}$, $ACC_{\mathcal{D}_{t}, \mathcal{D}_{t-i}}$ is the accuracy of the model trained on current domain  $\mathcal{D}_{t}$ and evaluated on previous domain  $\mathcal{D}_{t-i}$. For multi-label classification, Fr is calculated based on $l\omega lrap$ instead of accuracy. 
A higher average accuracy / $l\omega lrap$ and lower Fr are better. 
\begin{table}[]
 \caption{Average ASC accuracy of the methods over current and previously seen domains in domain-aware setup.}
     \centering
   \begin{tabular}{l|ccccc}
 \toprule
 & \multicolumn{5}{c}{\textbf{European cities (10 classes) $\rightarrow$ Korea (4 classes)}}\\
 Method & \makecell{$\mathcal{D}_1$\\6 cities} & \makecell{$\mathcal{D}_2$\\Lisbon} & \makecell{$\mathcal{D}_3$\\Lyon} & \makecell{$\mathcal{D}_4$\\Prague} & \makecell{$\mathcal{D}_5$\\Korea} \\
 \midrule
 FE & 67.3 & 52.9 &49.6 &41.3  & 22.7 \\
 FT & 67.3 & 52.1 &51.8 &41.4  & 31.1 \\
\midrule
Single-task & 67.3 & 53.1 &51.6 &38.7  & 34.1 \\
Multi-task & {67.3} & {64.1} &{66.2} & 67.0 &67.9 \\
\midrule
OD \cite{mulimani2023incremental} & 67.3 & 52.3 &54.2 &38.8  & 31.5 \\
ODFD \cite{mulimani2024class} & 67.3 & 53.1 &54.9 &41.4  & 35.6 \\
 
BN statistics \cite{mulimani2024online} & 67.3 & 57.7 &57.4 & 55.0&52.2 \\
 \midrule
 clf & 67.3 & 58.9 & 63.1 & 57.4 & 59.6 \\
 BN & 67.3 & 68.0 & 68.3 & 64.3 & 67.9 \\
 BN-clf & 67.3 & 68.4 & 69.4 & 65.9 & 69.4 \\
 ADIL & 67.3 & \textbf{69.6} &\textbf{71.4} & \textbf{68.7} & \textbf{71.9}\\
\midrule
\midrule
& \multicolumn{5}{c}{\textbf{Korea (4 classes) $\rightarrow$ European cities (4 classes)}}\\
 Method& \makecell{$\mathcal{D}_1$\\Korea} &\makecell{$\mathcal{D}_2$\\6 cities} & \makecell{$\mathcal{D}_3$\\Lisbon} & \makecell{$\mathcal{D}_4$\\Lyon} & \makecell{$\mathcal{D}_5$\\Prague}  \\
 \midrule
FE &  86.0 & 62.1 &59.4 &62.6  & 65.2\\
 FT &  86.0 & 61.8 &62.9 &68.5  & 65.3\\
\midrule
Single-task &  86.0 & 62.4 &61.1 &63.4  & 65.2\\
Multi-task &  86.0 & 86.2 &83.3 &88.0  & 87.4\\
\midrule
OD \cite{mulimani2023incremental} &  86.0 & 61.9 &62.1 &69.8  & 68.1\\
ODFD \cite{mulimani2024class} &  86.0 & 65.4 &62.9 &72.1  & 70.1\\
 
BN statistics \cite{mulimani2024online} & 86.0 & 65.6 &62.8 &58.0  & 58.3\\
 \midrule
 clf &  86.0 & 68.4 &60.4 &64.8  & 68.7\\
 BN & 86.0 & 84.1 &79.1 &80.4  & 80.0\\
 BN-clf &  86.0 & 84.2 &81.5 &82.6  & 82.2\\
 ADIL &  {86.0} & \textbf{84.4} &\textbf{82.4} &\textbf{83.8}  & \textbf{83.4}\\
 \bottomrule

 \end{tabular}     
     \label{tab:Aware}
     \end{table}
\subsection{Results of domain-aware setup}

\begin{figure*}[htbp!]
  \centering
  \subfloat[FE]{\includegraphics[width=5.85cm, height=2.7cm]{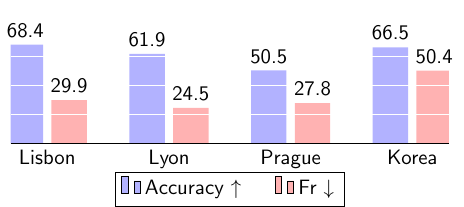}\label{Fig:FE}}
  \hfill
  \subfloat[FT]{\includegraphics[width=5.85cm, height=3cm]{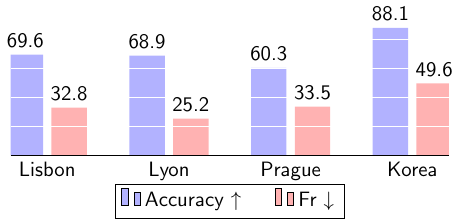}\label{Fig:FT}}
  \hfill
  \subfloat[Single-task]{\includegraphics[width=5.85cm, height=3cm]{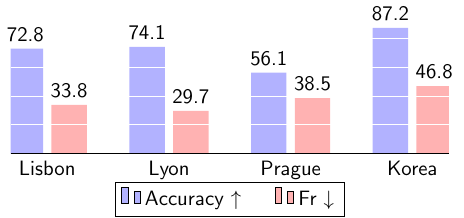}\label{Fig:disj}}   
  \caption{Accuracy at the current domain and average forgetting over previous domains of FE (a), FT (b) and single-task (c) methods for Europe $\rightarrow$ Korea.}
        \label{Fig:BASELINES}
     \vspace{-10pt}
\end{figure*}

\textbf{Acoustic scene classification:} We show the performance of the baselines and our proposed approach in Table \ref{tab:Aware}.  The base model is trained on domain $\mathcal{D}_1$ for an accuracy of 67.3\% when $\mathcal{D}_1$ is 6 European cities and 86.0\% when $\mathcal{D}_1$ is Korea. For a detailed analysis, the accuracy of the baseline systems on a current domain and forgetting of previously seen domains are compared in Fig.~\ref{Fig:BASELINES}.

The accuracy of FT and FE systems on the current domain in Fig.~\ref{Fig:FE} and ~\ref{Fig:FT} shows that allowing the feature extractor component, i.e., FT, to adapt to a new domain helps the model to achieve maximum plasticity rather than freezing it, i.e. FE. However, both FT and FE suffer from forgetting of previous domains due to domain shift, and exhibit minimum stability. There is high domain shift when going from Europe to Korea or Korea to Europe, leading to overwriting the knowledge of previously seen European/Korean cities. Hence, FE and FT suffer from higher forgetting. We observe lower forgetting within the European cities and also changed the order of European cities but did not observe any notable change.

The "single-task" baseline, trained from scratch on each location separately, is better than FT for Lisbon and Lyon, which can also be seen in Fig.~\ref{Fig:disj} and ~\ref{Fig:FT}. On the other hand, previous knowledge of FT improves the performance on Prague. The proposed ADIL combines both previous domain-shared and domain-specific characteristics and achieves comparable plasticity, i.e., accuracy on the current domain, with FT and single-task baselines, without forgetting any of the previously seen domains in all domain shift conditions. 

We also compare the performance of ADIL with state-of-the-art knowledge distillation (KD) methods reported in the literature for CIL setups. OD (output discrepancy) reduces the discrepancy in prediction space of the current model of $\mathcal{D}_t$ using  the previous model trained on $\mathcal{D}_{t-1}$ \cite{mulimani2023incremental}. ODFD (FD: feature discrepancy) reduces the discrepancy in both feature and prediction spaces \cite{mulimani2024class}. These distillation methods seem to work well to classify the sounds within a domain, but are less effective with domain shift, resulting in lower average accuracy. 

\begin{figure}[!tbp]
  \centering
  \includegraphics[width=0.4\textwidth, height=3cm]{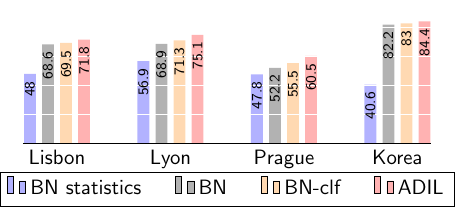}  
  \caption{Accuracy of the BN-based methods: BN statistics, BN, BN-clf and  ADIL at different domains in the order of Europe $\rightarrow$ Korea}
     \label{fig:ADIL}
     \vspace{-10pt}
\end{figure}

Our previous work on DIL \cite{mulimani2024online}, which only corrects the statistics of the BN layers ( BN statistics in Table~\ref{tab:Aware}), and avoids the forgetting. However, this method poorly adapts to the new domains and exhibits lower plasticity. The accuracy of BN statistics on each current domain and average performance over all learned domains are given Fig.~\ref{fig:ADIL} and Table \ref{tab:Aware} respectively.

We also compare the contribution of each component of our network architecture in Table \ref{tab:Aware} and Fig.~\ref{fig:ADIL}: (1) The clf system uses a separate classifier for each domain, with all other parameters frozen. It is a multi-head FE and exhibits poor performance without domain-specific BN layers. (2) The BN system uses domain-specific BN layers, with a common base classifier shared by all domains. It outperforms all the baselines and improves the plasticity significantly as compared to the BN statistics system. (3) To further improve the performance of BN, we add a separate classifier for each domain, resulting in BN-clf system. (4) We found that combining the output of each domain-specific classifier to a base classifier, resulting in ADIL, which improves the overall performance. For instance, the ADIL improves the performance of Prague by 5\%p (percentage point) as compared to BN-clf.  This is because ADIL uses both previous domain-shared and domain-specific knowledge to classify acoustic scenes, and achieves a higher performance on any new domain as compared to any of the BN-based systems. We observed similar behavior in the Korea $\rightarrow$ Europe setup.

\textbf{Multi-label audio classification:} 
We compare the performance of the baseline systems: FE, FT and single-task with our proposed ADIL system for both Audioset $\rightarrow$ FSD50K and FSD50K $\rightarrow$ Audioset in Table \ref{tab:Audioset}. The base model is trained on domain  $\mathcal{D}_1$ and achieved an  $l\omega lrap$ of 49.7\% on Audioset and 34.1\% on FSD50K.
One can observe the lower performance of single-task when going from Audioset $\rightarrow$ FSD50K. Training the model from scratch on FSD50K gives lower $l\omega lrap$ compared to Audioset, significantly forgetting previous Audioset classes due to domain shift and achieving lower average $l\omega lrap$. FT and FE systems benefit from the previous knowledge of Audioset and effectively adapt to FSD50K, but also suffer from forgetting. The proposed ADIL  effectively adapts well to the FSD50K by learning domain-specific parameters, outperforming all other  systems.

Going from FSD50K $\rightarrow$ Audioset, the $l\omega lrap$ of a single-task baseline trained on Audioset with 35 classes is 28.7\%; this reduced performance compared to the earlier 50 classes may be due to lesser data. Fine-tuning the model trained on FSD50K to Audioset, i.e., FT, did not improve the performance of the model on Audioset. Single-task and FT also show higher forgetting, resulting into reduced average $l\omega lrap$. However, FE only updates the classifier and improves the $l\omega lrap$ on Audioset to 34.2\% with reduced forgetting. Our approach ADIL, updates both domain-specific classifier and BN layers, and further improves  $l\omega lrap$ on Audioset to 47.2\% without forgetting any of the previous classes from FSD50K. The results show the benefit of the proposed ADIL approach which works in all domain-shift conditions where other methods fail.

\begin{table}[]
 \caption{Average $l\omega lrap$ of the methods over current and all previously seen domains in domain-aware setup. The value in parenthesis indicates forgetting}
     \centering
   \begin{tabular}{l|cc|cc}
 \toprule
 Method & \makecell{$\mathcal{D}_1$ (50)\\AudioSet} & \makecell{$\mathcal{D}_2$ (35)\\FSD50K}  & \makecell{$\mathcal{D}_1$ (35)\\FSD50K} &\makecell{$\mathcal{D}_2$ (35)\\AudioSet} \\
  \midrule
  FE & 49.7 & 44.9 \textcolor{red}{(15.6)} &34.1 &32.4 \textcolor{red}{(3.6)}   \\
 FT & 49.7 & 45.8 \textcolor{red}{(18.1)} &34.1 &26.4 \textcolor{red}{(9.2)}\\
\makecell{Single-task} & 49.7 & 22.8 \textcolor{red}{(34.0)} &34.1 &26.8  \textcolor{red}{(9.3)} \\

\midrule

 ADIL & 49.7 & \textbf{47.5} &{34.1} & \textbf{40.7} \\
 \bottomrule
 \end{tabular}     
     \label{tab:Audioset}
       \end{table}

\subsection{Results of domain-agnostic setup}
Results of the proposed ADIL approach in domain-agnostic setup are  given in Table \ref{tab:Agnostic}. The performance of our approach depends on the accurate prediction of the domain-specific layers using Eq. (\ref{uncertainity}). Unlike domain-aware setup, here ADIL is affected by forgetting due to the incorrect selection of domain-specific layers. However, ADIL balances the stability-plasticity trade-off at considerable level and still outperforms all compared  baselines from Table \ref{tab:Aware} on ASC. In the more challenging, imbalanced, multi-label audio classification, the performance of ADIL is better than single-task baseline, which is trained on a single domain.

\begin{table}[]
 \caption{Average accuracy and $l\omega lrap$ of the methods over current and all previously seen domains in domain-agnostic setup.}
     \centering
   \begin{tabular}{l|c|c|c|c|c}
 \toprule
 Order & \makecell{$\mathcal{D}_1$} & \makecell{$\mathcal{D}_2$} & \makecell{$\mathcal{D}_3$} & \makecell{$\mathcal{D}_4$} &  \makecell{$\mathcal{D}_5$}\\
 \midrule
 Europe$\rightarrow$Korea & 67.3 & 61.9 &62.8 &59.4 &53.0 \\
Korea$\rightarrow$Europe & 86.0 & 69.3 &71.9 &72.6  & 73.7\\
 Audioset$\rightarrow$FSD50K & 49.7 & 32.0 & &  & \\
FSD50K$\rightarrow$Audioset & 34.1 & 29.4 & & &\\ 
 \bottomrule
  \end{tabular}     
     \label{tab:Agnostic}
         \end{table}   
\section{Conclusion}
In this paper, we presented  
the ADIL approach to solve single and multi-label audio classification problems over time. The proposed approach shares 84\% of the parameters of the base model among all domains and only needs to train domain-specific parameters in each incremental time step. Our approach outperforms all the baselines when domain-shift is present in domain-aware setup, and also shows promising results in domain-agnostic setup. 
Improving the performance of the ADIL in domain-agnostic setup is one of our future research problems.
\bibliographystyle{IEEEtran}
\bibliography{references}
\end{document}